\documentclass{elsart}
\usepackage{epsfig}
\usepackage{amssymb}

\begin{document}

\def\simlt{\mathrel{\rlap{\lower 3pt\hbox{$\sim$}}\raise 2.0pt\hbox{$<$}}}
\def\simgt{\mathrel{\rlap{\lower 3pt\hbox{$\sim$}} \raise 2.0pt\hbox{$>$}}}
\def\di{\mbox{d}}
\def\Msun{M_{\odot}}
\def\HI{\hbox{H$\scriptstyle\rm I\ $}}
\def\etal{{\it et al.~}}

\newcommand{\q}{\begin{equation}}
\newcommand{\qa}{\begin{eqnarray}}
\newcommand{\qs}{\begin{eqnarray*}}
\newcommand{\nq}{\end{equation}}
\newcommand{\nqa}{\end{eqnarray}}
\newcommand{\nqs}{\end{eqnarray*}}
\newcommand{\ud}{\mathrm{d}}
\def\astrobj#1{#1}

\begin{frontmatter} 
\title{Gamma-ray constraints on the infrared background excess} 
\author[SISSA]{M. Mapelli},
\author[COMO]{R. Salvaterra},
\author[SISSA]{A. Ferrara} 

\address[SISSA]{SISSA/International School for Advanced Studies, Via Beirut 4, 34014 Trieste, Italy}
\address[COMO]{Universit\`a dell'Insubria, Via Valleggio 11, 22100 Como, Italy}
 
\maketitle \vspace {4cm }
 
\begin{abstract}
Motivated by the idea that the recently detected near-infrared (1.2-4 $\mu$m) excess over the contribution of known galaxies is due to redshifted light from the first cosmic stars (Salvaterra \& Ferrara 2003), we have used the effect caused by photon-photon absorption on gamma-ray spectra of blazars to put constraints on extragalactic background light (EBL) from the optical to the far-IR bands. Our analysis is mainly based on the blazar \astrobj{H~1426+428}, for which we assume a power-law unabsorbed spectrum. 
We find that an EBL model with no excess over known galaxies in the near-infrared background (NIRB) is in agreement with all the considered blazars; however, it implies a very peculiar intrinsic spectrum for \astrobj{H~1426+428}. 
Additional data on the blazars \astrobj{1ES1101-232}, \astrobj{H~2356-309} and \astrobj{PKS~2155-304} exclude the existence of a strong NIRB excess consistent with Kelsall's model of zodiacal light subtraction (ZL); the COBE/DIRBE measurements, after Wright's model ZL subtraction, represent a firm NIRB upper limit.
The constraints on the optical EBL are weaker, due to the fact that predictions from different optical EBL models are often comparable to the experimental errors. 
In the mid-infrared the {\it SPITZER} measurement of $\nu{}I_{\nu{}}$=2.7 nW m$^{-2}$ sr$^{-1}$ at 24 $\mu{}$m, gives a good fit for all the considered blazars.
\end{abstract}
\begin{keyword}
gamma rays: theory - infrared: general - quasars: general - quasars: individual: \astrobj{H~1426+428} 

\end{keyword}

\end{frontmatter}

\section{Introduction}

The study of the  Extragalactic Background Light (hereafter EBL) might provide unique information to understand many crucial astrophysical questions, 
including, among others, the first cosmic star formation (Salvaterra \&{} Ferrara 2003), the evolution of  galaxies (Totani \&{} Takeuchi 2002; 
Totani et al. 2001), the role of  dust emission (Granato et al. 2001).
Yet, the EBL determination in the optical and infrared bands poses challenging observational and theoretical difficulties, often leading to 
discrepant interpretations. For example, Madau \& Pozzetti (2000) suggest that the optical EBL can be completely explained by galactic counts, 
while Bernstein et al. (2002) claim the presence of an optical excess. 
Moreover, the presence of an excess (Matsumoto et al. 2000; Wright 2001) in the near infrared background (hereafter NIRB) crucially depends on 
the subtraction of the zodiacal light (Kelsall et al. 1998; Wright 1997, 1998; Wright \& Reese 2000). 
For the middle and far infrared background (hereafter MIRB and FIRB) various models have been proposed (Totani \&{} Takeuchi 2002; Silva et al. 2004; 
Primack et al. 1999) which cannot be discriminated on the basis of currently available data (Elbaz et al. 2002; Metcalfe et al. 2003, and references here).
Finally, very strong constraints to the MIRB can be provided by the {\it SPITZER}\footnote{http://www.spitzer.caltech.edu/index.shtml} satellite 
(Papovich et al. 2004), which indicates a background total flux of $2.7^{+1.1}_{-0.8}$ nW m$^2$ sr$^-1$ at 24 $\mu$m. 
In the next years {\it SPITZER} (and other projects like ASTRO-F and SPICA) will likely be able to reconstruct the complete map of the MIRB and FIRB. \\
Waiting for the upcoming infrared facilities, an alternative way to put independent constraints to the EBL is to resort to the so-called 
photon-photon absorption of the high energy tail of the blazar spectra. It is known that gamma rays in the GeV-TeV energy bands can be absorbed 
by softer (mainly optical and infrared) photons, via electron-positron pair production (Nikishov 1962). 
Stecker, de Jager \& Salamon (1992) pioneered the method on the spectrum of 3C 279 to first study the EBL. Their attempt stimulated 
a plethora of works on the subject  (Stecker, de Jager \& Salamon 1993; Madau \& Phinney 1996; Stecker \& de Jager 1997; Aharonian et al. 2003; 
Konopelko et al. 2003; Stecker 2003; Costamante et al. 2004; Dwek \& Krennrich 2005) that were encouraged by the availability of new measurements of blazar 
spectra in the TeV regime based on the Cherenkov imaging technique (Aharonian et al. 1999b; Krennrich et al. 1999 and Aharonian et al. 2002b for \astrobj{Mkn 421};  Aharonian et al. 1999a, 2001a, 2002b for \astrobj{Mkn 501}; Aharonian et al. 2003 for \astrobj{H~1426+428}).

In this paper we calculate the optical depth due to photon-photon absorption using the most recent optical and infrared background measurements 
(Bernstein et al. 2002; Matsumoto et al. 2000; Papovich et al. 2004), combined with state-of-art theoretical models of the NIRB 
(Salvaterra \& Ferrara 2003) and of the MIRB-FIRB (Totani \& Takeuchi 2002). We compare these results with the Cherenkov imaging data available 
for six blazars (\astrobj{H~1426+428}, \astrobj{Mkn 421}, \astrobj{Mkn 501}, \astrobj{PKS~2155-304}, \astrobj{1ES1101-232}, \astrobj{H~2356-309}), to derive some constraints on the EBL. The final aim of the study is to assess if the most recent 
EBL data and models exclude or require the presence of an excess with respect to galaxy counts in the optical and in the near-infrared regime. In addition,   
we try to test the predictions of the proposed EBL models, especially in the optical, near and middle infrared ranges. \\
The paper is organized as follows. Sec. 2 presents the method adopted to derive the optical depth for the 
photon-photon absorption and to compare it with blazar data. 
In Sec. 3 we briefly review the most important EBL data and the EBL models considered. 
Sec. 4 is devoted to the description of  the Cherenkov data available for the 
three blazars that we analyze. In Sec. 5 we discuss the results and Sec. 6 summarizes our conclusions.

\section{The photon-photon absorption}
The optical depth due to photon-photon absorption can be written (Stecker et al. 1992; Madau \& Phinney 1996) as
\q\label{eq:eq1}
\tau{}(E)=\int^{z_{em}}_{0}d{z}\frac{d{l}}{d{z}}\int^{1}_{-1}d{x}\,{}\frac{(1-x)}{2}\int^{\infty{}}_{\epsilon{}_{th}}d{\epsilon{}}\,{}n(\epsilon{})\,{}\sigma{}(\epsilon{},E,x), 
\nq
where $d{l}/d{z}$ is the proper line element\footnote{We adopt the following cosmological parameters: 
Hubble constant $H_0$=71 km s$^{-1}$ Mpc$^{-1}$, $\Omega{}_M$=0.27, 
$\Omega{}_\Lambda{}$=0.73, which are in agreement with the recent WMAP determination (Spergel et al. 2003).}
\qs
\frac{d{l}}{d{z}}=\frac{c}{H_0}\left[(1+z)\,{}{\mathcal E}(z)\right]^{-1},
\nqs
$c$ is the speed of light and
\qs
{\mathcal E}(z)=\left[\Omega{}_M(1+z)^3+\Omega{}_\Lambda{}+(1-\Omega{}_\Lambda{}-\Omega{}_M)(1+z)^2\right]^{1/2}.
\nqs
In eq. \ref{eq:eq1}, $x\equiv{}\cos{\theta{}}$, $\theta{}$ being the angle between the directions of the two interacting photons. $E=E_0\,{}(1+z)$ is the observed energy of the blazar photon and $\epsilon{}=\epsilon{}_0(1+z)$ is the observed energy of the background photon; $z_{em}$ is the redshift of the considered blazar; finally,
$n(\epsilon)$ is the specific number density of background photons.
The energy threshold for the interaction, $\epsilon{}_{th}$, is defined by:
\q\label{eq:eq2}
\epsilon{}_{th}=\frac{2\,{}m_e^2\,{}c^4}{E\,{}(1-x)}
\nq
where $m_e$ is the electron mass.\\
The photon-photon absorption cross section (Heitler 1960; Gould \& Schr\'eder 1967) is given by
\q\label{eq:eq3}
\sigma{}(\epsilon{},E,x)=\frac{3\,{}\sigma{}_T}{16}(1-\beta{}^2)\left[2\,{}\beta{}(\beta{}^2-2)+(3-\beta{}^4)\ln{\left(\frac{1+\beta{}}{1-\beta{}}\right)}\right]
\nq
where $\sigma_T$ is the Thomson cross section and
\qs
\beta{}\equiv{}\left[1-\frac{2\,{}m_e^2\,{}c^4}{E\,{}\epsilon{}\,{}(1-x)\,{}}\right]^{1/2}
\nqs
The cross section $\sigma$ peaks sharply 
at $\lambda{}\sim{}hc\,{}E/(2m_e^2c^4) \sim{}
2.4 (E/{\rm TeV}) \mu$m (where $\lambda{}$ is the wavelength of 
the interacting background photon). 
Fig.~1 shows the behavior of $\sigma(\lambda)$
integrated over the angle $\theta{}$. Fig.~2 shows how the cross section depends on the interaction 
angle $\theta{}$. 

Substituting into eq. \ref{eq:eq1} the relation $n(\epsilon{})=n(\epsilon{}_0)(1+z)^3$ (Madau \& Phinney 1996), we finally obtain: 
\qa\label{eq:eq4}
\tau{}(E)=\frac{c}{2\,{}H_0}\int^{z_{em}}_{0}d{z}\frac{(1+z)^2}{{\mathcal E}(z)}\int^{1}_{-1}d{x}\,{}(1-x) 
\int^{\infty{}}_{\epsilon{}_{th0}}d{\epsilon{}_0}\,{}n(\epsilon{}_0)\,{}\sigma{}(\epsilon{}_0,E_0,x,z); 
\nqa
$n(\epsilon{}_0)$ is related to the observable quantity $F(\epsilon{}_0)$, the  background photon flux at redshift $z=0$, by the simple relation:
\q\label{eq:eq5}
n(\epsilon{}_0)=4\pi{}\,{}\frac{F(\epsilon{}_0)}{c\,{}\epsilon{}_0^2}\quad{}\textrm{cm}^{-3}\textrm{ erg}^{-1}
\nq 

\begin{figure}
\center{{
\epsfig{figure=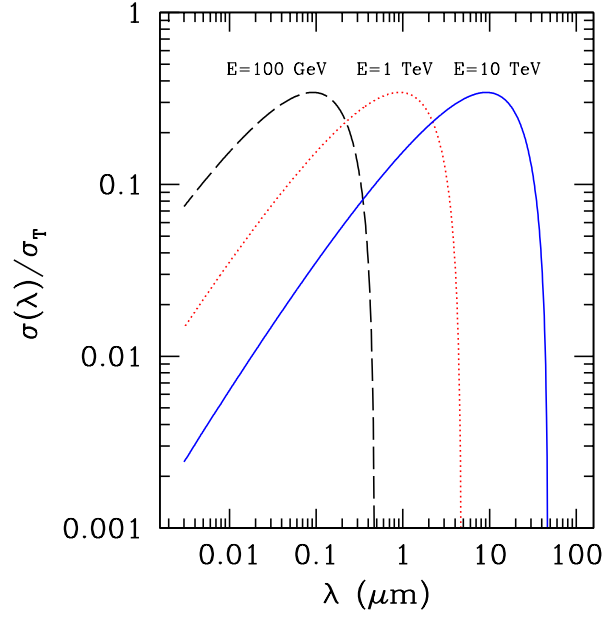,height=9cm}
}}
\caption{\label{fig:fig1} Photon-photon absorption cross section, $\sigma{}(\lambda{})$, as a function of the background photon wavelength, $\lambda$, integrated over the interaction angle, for three values of the observed energy of the blazar photon: 100 GeV ({\it dashed line}), 1 TeV ({\it dotted}) and 10 TeV ({\it solid}). $\sigma{}_T$= 6.652$\times{}$10$^{-25}$ cm$^2$ is the Thomson cross section.} 
\end{figure}

\begin{figure}
\center{{
\epsfig{figure=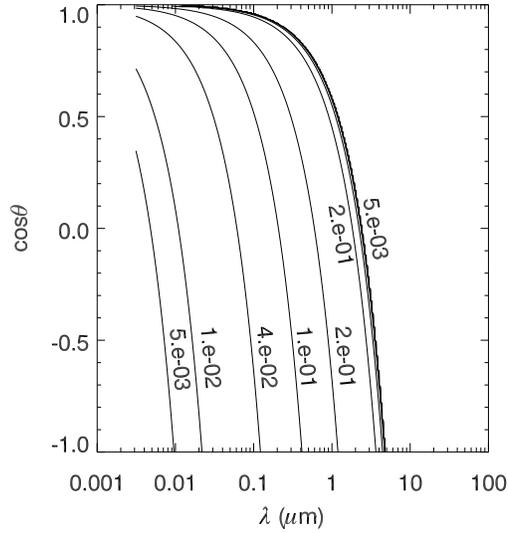,height=8cm}
}}
\caption{\label{fig:fig2} Isocontours  of the cross section  (in units of $\sigma{}_T$) as a function of $\theta{}$ and $\lambda{}$ for $E$=1 TeV.
}
\end{figure}

\bigskip
We have calculated numerically $\tau{}(E)$ using a three-dimensional integral based on the method of Gaussian quadratures (Press et al. 1992). The integration accuracy varies between 
$3\times{}10^{-5}$\% and $8\times{}10^{-3}$\% in the range 0.1-10 TeV.
$\tau{}(E)$ has been estimated for different values of $F(\epsilon{}_0)$, whose
choice has been guided by both observational data and theoretical models (see next Section). To discriminate among these different estimates of $\tau{}(E)$ we need a comparison with observational data of blazar spectra, which we have performed in two independent ways:\\
\begin{enumerate}
\item The first method  assumes a theoretical model of the unabsorbed blazar spectrum. In particular, we adopt a simple power-law spectrum \\$(d{N}/d{E})_{unabsorbed}\propto{}E^{-\alpha{}}$ for the blazar \astrobj{H~1426+428}, the main target of our analysis. For the other blazars considered  (\astrobj{Mkn~421} and \astrobj{Mkn~501}) the unabsorbed spectrum is better fitted by a power-law with an exponential cut-off (Konopelko et al. 2003). Some theoretical models suggest that the blazar spectrum must present a cut-off in the TeV range (Inoue and Takahara 1996; Tavecchio, Maraschi \& Ghisellini 1998; Fossati et al. 2000). 

The absorbed spectrum is then obtained by convolving the unabsorbed spectrum with
$\tau(E)$,  
\q\label{eq:eq6}
\left(\frac{d{N}}{d{E}}\right)_{absorbed}=e^{-\tau{}(E)}\left(\frac{d{N}}{d{E}}\right)_{unabsorbed},
\nq
and by changing the spectral index $\alpha$ to obtain 
the best fit to the observed blazar spectrum.              
\item The alternate method consists in inverting equation \ref{eq:eq6}.
In this case we apply $\tau{}(E)$ directly to the observational data, to derive the 
unabsorbed spectrum and check if the intrinsic spectral index $\alpha$ is consistent 
with the current theoretical predictions. This procedure does not require any {\it a priori} assumption about the unabsorbed spectrum shape. 
An important sanity check is to verify whether the values of $\alpha$ derived from the two methods are consistent. 
\end{enumerate}

\section{The optical-infrared background}
\subsection{Observations of the EBL}
The major difficulty in measuring the EBL arises from the subtraction of the 
interplanetary dust scattered sunlight (i.e. zodiacal light, ZL) contribution. 
This problem presents different aspects and complexity according to the range of wavelengths observed. A vast amount of literature is present on the subject (see
Hauser \& Dwek 2001 for a complete review); here, we only concentrate on the the EBL 
measurements and models used in our work.

\subsubsection{Galaxy counts}
Deep optical and near infrared galaxy counts give an estimate of the EBL fraction
coming from normal galaxies. Madau \& Pozzetti (2000) derived the contribution of 
known galaxies in the $UBVIJHK$ bands from the $Southern\,{}Hubble\,{}$\\$Deep\,{}Field$ 
imaging survey. In particular for the U, V, B and I bands (corresponding to the wavelengths $\lambda{}$ = 3600, 4500, 6700 and 8100 \AA) they found a mean flux respectively $2.87_{-0.42}^{+0.58}$, $4.57_{-0.47}^{+0.73}$, $6.74_{-0.94}^{+1.25}$ and $8.04_{-0.92}^{+1.62}$ in units of  10$^{-6}$ erg s$^{-1}$ cm$^{-2}$ sr$^{-1}$.

\subsubsection{Optical excess}
Estimates of the optical EBL based on photometric scans across dark nebulae (Mattila 1976; Spinrad \&{} Stone 1979) and on photoelectric measurements (Dube et al. 1977, 1979) provide upper limits that are higher than the flux given by galaxy counts alone at the same wavelengths. In particular the most recent work by Dube et al. (1979) provides an upper limit of $2.6\times{}10^{-5}$ erg s$^{-1}$ cm$^{-2}$ sr$^{-1}$ at $\lambda{}$= 5115 \AA.
Bernstein et al. (2002) measured the mean flux of the optical EBL at 3000, 5500, and 8000 \AA, using the Wide Field Planetary Camera 2 (WFPC2) and the Faint Object Spectrograph, both on board the {\it Hubble Space Telescope}, combined with the du Pont 2.5 m Telescope at the Las Campanas Observatory. They found for these three band a mean flux of the EBL respectively $12.0_{-6.3}^{+17.7}$, $14.9_{-10.5}^{+19.3}$, and $17.6_{-12.8}^{+22.4}$ in units of 10$^{-6}$ erg s$^{-1}$ cm$^{-2}$ sr$^{-1}$, considerably higher than the contribution of the galaxy counts alone. 
In this band not only ZL is likely to provide a
substantial contribution, but also terrestrial airglow, and dust-scattered Galactic 
starlight (diffuse Galactic light) might represent a potential problem for the
measurement. The impact of such systematic errors has led Mattila (2003) to
question the claim by Bernstein et al. of the discovery of an optical EBL
excess.

\subsubsection{Near Infrared Background: DIRBE and NIRS data}
The available NIRB data come from the Diffuse Infrared Background Experiment (DIRBE) on board of the Cosmic Background Explorer (COBE) and from the Near InfraRed Spectrometer (NIRS) on board of the InfraRed Telescope in Space (IRTS). The DIRBE instrument provided a survey of the sky in 10 photometric bands at 1.25, 2.2, 3.5, 4.9, 12, 25, 60, 100, 140, and 240 $\mu{}m$ using a 0.7$^{\circ{}}\times{}$0.7$^{\circ{}}$ field of view. A summary of the DIRBE results can
be found in Hauser et al. (1998). The NIRS instrument covers the wavelength range from 1.4 to 4.0 $\mu{}$m with a spectral resolution of 0.13 $\mu{}$m. Matsumoto et al. (2000) made a preliminary analysis of the NIRS data, estimating the NIRB on the basis of the 5 NIRS observation days unperturbed by atmospheric, lunar and nuclear radiation effects.

Both the DIRBE and the NIRS data show an excess in the NIRB with respect to galaxy counts. An estimate of this excess depends on a critical point: the subtraction of the contribution of the ZL from the measurements.
There are at least two models of ZL. The model described in Kelsall et al. (1998) exploits the temporal variability of the signal caused 
by looking at the sky through different amounts of the interplanetary dust as the
Earth orbits the Sun. Wright \& Reese (2000) noted that the high value of the
EBL flux derived by Kelsall from the DIRBE data at 25 $\mu{}$m might indicate
that a residual ZL flux could remain after subtraction.
Thus they suggest a different approach, which requires that the EBL signal  
at 25 $\mu{}$m after ZL subtraction is zero (Wright 1997, 1998).
In practice, subtracting the ZL both with the Kelsall and with the Wright method the presence of a NIRB excess is unquestionable, even if the amount of this excess depends on the ZL model assumed.

\subsubsection{Mid Infrared Background: {\it SPITZER} data}
Before the {\it SPITZER} satellite, the only MIRB available data came from the ISOCAM 
deep extragalactic surveys. From the analysis of the ISOCAM number counts Elbaz et 
al. (2002) computed an EBL flux (integrated down to 50 $\mu{}$Jy) of $2.4\pm{}0.5$ nW m$^{-2}$ sr$^{-1}$ at 15$\mu{}$m 
(68$\%{}$ confidence level). In deriving this value, Elbaz et al. took into account, among other 
surveys, of 15 $\mu$m counts from a portion of the ISO gravitational 
lensing survey. Metcalfe et al. (2003) use a full lensing survey (covering 
Abell 2218, Abell 2390 and Abell 370), and, by integrating from 30 $\mu{}$Jy 
up-wards, obtain an EBL flux of $2.7\pm{}0.62$ nW m$^{-2}$ sr$^{-1}$ at 15$\mu{}$m (68$\%{}$ confidence level). This result is consistent with the upper limit of 5 nW m$^{-2}$ sr$^{-1}$ on the 15 $\mu{}$m EBL, estimated by Stanev \&{} Franceschini (1998). This upper limit is calculated from photon-photon absorption effects on the spectra of \astrobj{Mkn~501}; note, though, that this analysis is based on relatively poor-quality, old data (Aharonian et al. 1997). 

Recently, integrating to 60 $\mu{}$Jy the counts from the Multiband Imaging Photometer on board of the {\it SPITZER} satellite (MIPS, Rieke et al. 2004), Papovich et al. (2004) found a lower limit to the EBL flux at 24$\mu$m of $1.9\pm{}0.6$ nW m$^{-2}$ sr$^{-1}$. Extrapolating to fainter flux densities, they derive an estimate of the total 24 $\mu{}$m background of $2.7^{+1.1}_{-0.7}$ nW m$^{-2}$ sr$^{-1}$, in good agreement with the result of Metcalfe et al. (2003). Other lower limits to the EBL flux at wavelengths ranging from 3 to 10 $\mu{}$m, derived by {\it SPITZER} measurements (Fazio et al. 2004), are reported in Fig.~3.

\subsection{Theoretical models of the EBL}
The somewhat sparse experimental information available on the  extended
wavelength range of the EBL forces us to resort to theoretical modeling in
order to fully reconstruct its spectrum. In the following we summarize the
main features of the EBL models used in this work.

\subsubsection{Contribution of  Pop~III stars to the NIRB}
The most plausible explanation of the NIRB excess 
 is that the EBL in this wavelength range is due to the redshifted UV and optical light emitted by Pop~III stars (Bond et al. 1986; Santos et al. 2002; Salvaterra \&{} Ferrara 2003). In particular Salvaterra \&{} Ferrara (2003) developed a model of the NIRB which, accounting for the most recent predictions of  Pop~III stellar spectra (Schaerer 2002) and IMF, nebular emission (i.e. the radiation coming from the nebula surrounding the star), and L$_{y\alpha{}}$ photons scattered by the intergalactic medium, is able to fit the NIRS data (Matsumoto et al. 2000) and the DIRBE data with both the methods of ZL subtraction. Their best fit predicts, for a star formation efficiency $f_{\ast{}}=0.1-0.5$, depending on the adopted IMF, a transition from (very massive)
Pop~III to Pop~II stars occurring at $z\approx 9$. This model is supported also by the analysis of the Infrared Background fluctuations performed by Magliocchetti et al. (2003). Such interpretation of the NIRB in terms of Pop~III stars, although very intriguing, might be somewhat extreme in terms of
the high star formation efficiencies and production of intermediate mass black holes, as pointed out also by Madau \& Silk (2005). \\
To test this model and
to understand the physical nature of the transition, a better knowledge of the 
EBL at $\lambda < 1.2\,{} \mu{}$m is crucial (i.e. a confirmation or rejection of the optical excess measured by Bernstein et al.). 

\subsubsection{Mid- and Far Infrared Background}
For the middle and the far infrared background a plethora of different models is
present in the literature. Among the most complete ones, that by Totani \&{}
Takeuchi (2002) is also consistent with the ISOCAM and the {\it SPITZER} measurements. These
authors construct a model of the near, middle and far infrared background based  on the backward approach for the luminosity evolution of galaxies (i.e. they infer 
the star formation history from the present-day galaxy optical-IR SEDs 
and chemical properties). Their model is characterized by a realistic treatment of 
dust and a physical determination of its temperature.

For our work we have considered also other models of the IR background, such as 
those presented by Primack et al. (1999) or by Silva et al. (2004). The results obtained using these models,
however, do not give as good fits to both the Spitzer and TeV data as that of Totani \& Takeuchi.

\subsection{Summary of the adopted EBL model}

\begin{table}
\begin{center}
\caption{Summary of the considered EBL models. 
}
\begin{tabular}{lll}
\hline
\hline
{\bf Model} & Optical Background & NIRB \\
\hline
C1          & \multicolumn{2}{c}{Madau \& Pozzetti (2000)}  \\
MK1          & Madau \& Pozzetti (2000)  & Matsumoto et al. (2000) (K)$^{a}$\\
MK2          & Bernstein et al. (2002)  & Matsumoto et al. (2000) (K)\\
DW1          & Madau \& Pozzetti (2000) & Wright (2001) (W)$^{a}$ \\
DW2          & Bernstein et al. (2002) & Wright (2001) (W) \\
\hline
\end{tabular}
\end{center}
\begin{flushleft}
{\footnotesize $^{a}$(K) and (W) indicate the ZL subtraction obtained using Kelsall's model and Wright's model, respectively. For the MIRB and FIRB we adopted always the  Totani \& Takeuchi (2002) model rescaled to the Spitzer data ( Papovich et al. 2004).}\\

\end{flushleft}
\label{tab_1}
\end{table}

The results of our study (presented in Sec. 5) are obtained adopting the
following assumptions on the EBL.
\begin{itemize}
\item For the optical background (i.e. between 0.3 and 1.2 $\mu{}$m) we have considered both the values obtained from galaxy counts only, following Madau \& Pozzetti (2000), and the case of a background excess (Bernstein et al. 2002 data, including their upper and lower limits). 
\item For the NIRB ($1.2 < \lambda <  4.\,{}\mu{}$m) we have used both 
the NIRS data (Matsumoto et al. 2000; Salvaterra \& Ferrara 2003) with the Kelsall's model of ZL subtraction, and  the DIRBE data with the Wright model of ZL subtraction (Gorjian et al. 2000; Wright \& Reese 2000; Wright 2001; Wright \& Johnson 2001).
\item For the MIRB and FIRB ($\lambda > 4.\,{}\mu{}$m)  we adopt the Totani \& Takeuchi (2002) model, rescaled by a factor 1.2 in the 8-30 $\mu{}$m range to match the {\it SPITZER} 24 $\mu$m
data point; in this wavelength range the uncertain contribution of spiral galaxies 
allows such rescaling. 
\end{itemize}
Finally we have considered a case in which the optical/NIRB in $0.3 < \lambda <
4\,{}\mu{}$m is contributed purely by galaxies as given by their counts (i.e. no excess) 
leaving the MIRB/FIRB as above. 
A summary of the considered models is given in Table~1. Fig.~3 shows these models and the considered data. 
\begin{figure*}
\center{{
\epsfig{figure=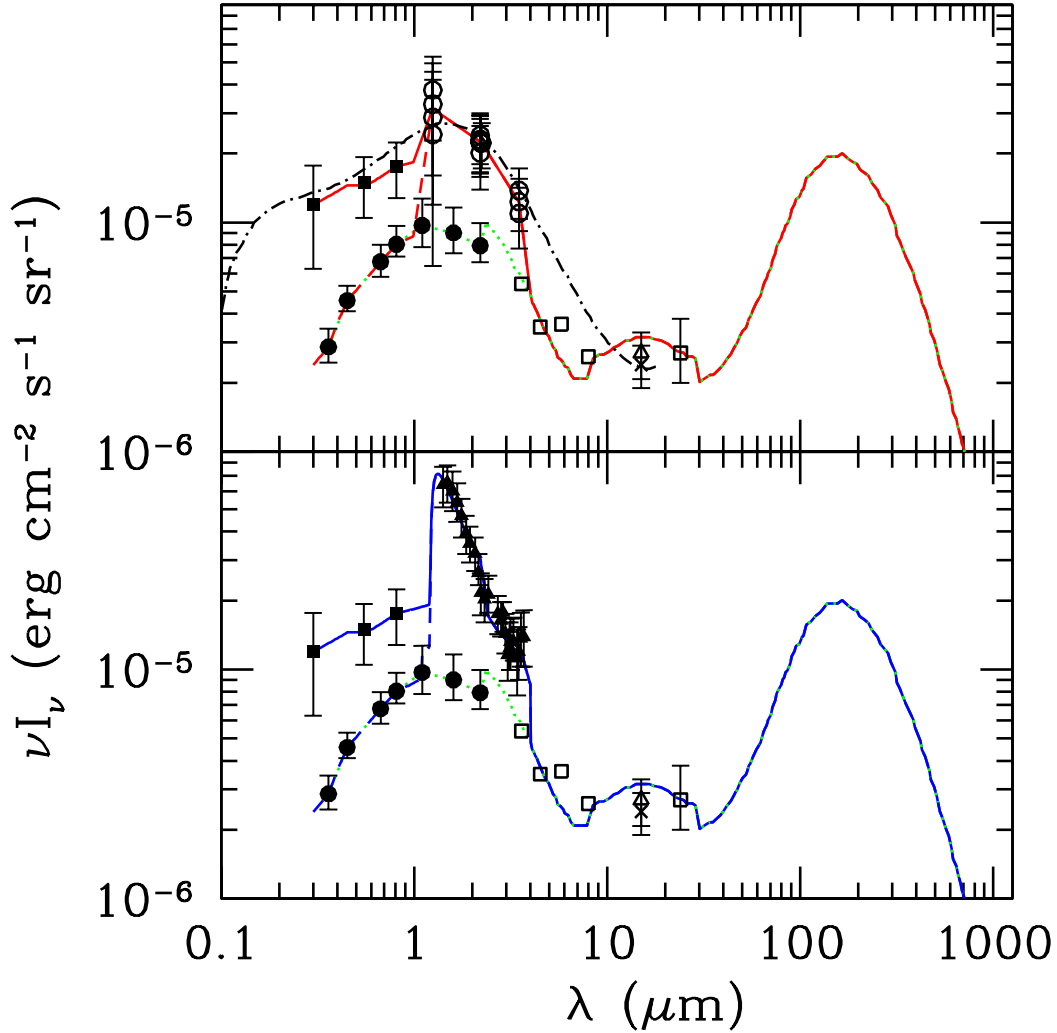,height=15cm}
}}
\caption{\label{fig:fig3} EBL data and corresponding models. In both the two panels are shown: data from Bernstein et al. ({\it filled squares}), Madau \& Pozzetti ({\it filled circles}), Elbaz et al. ({\it cross}), Metcalfe et al. ({\it open triangle}), Fazio et al. ({\it open squares} from 3 to 10 $\mu{}$m) and Papovich et al. ({\it open square} at 24 $\mu{}$m). The {\it dotted line} indicates C1 model. {\bf Upper panel:}  DW1 ({\it dashed line}) and DW2 ({\it solid line}) models. The DW data (Gorjian et al. 2000; Wright \& Reese 2000; Wright 2001; Wright \& Johnson 2001) are represented with {\it open circles}. For comparison, the {\it dot-dashed line} shows the P1.0 model of Aharonian et al. 2005b.  { \bf Lower panel:} MK1 ({\it dashed line}) and MK2 ({\it solid line}) models. The MK data (Matsumoto et al. 2000) are represented with {\it filled triangles}.}
\end{figure*}
In our work we neglect a possible redshift evolution of the EBL. This assumption is reasonable given the low redshift of the considered blazars, including the most distant one ($z=0.129$). We also neglect the possible self-absorption of the blazar, i.e. the contribution to the photon-photon absorption given by infrared photons produced by the blazar, as it can be shown to be  irrelevant (Protheroe \& Biermann 1997).

\section{TeV blazar spectra}
Sufficiently high resolution blazar spectra in the TeV regime, obtained with imaging Cherenkov techniques, are available for at least three blazars: \astrobj{H~1426+428}, \astrobj{Mkn~421} and \astrobj{Mkn~501}. Data have become recently available for other three blazars: \astrobj{PKS~2155-304} (Aharonian et al. 2005a), \astrobj{1ES1101-232} and \astrobj{H~2356-309} (Aharonian et al. 2005b).
\astrobj{H~1426+428} is at relatively high redshift ($z=0.129$); its TeV spectrum determination is therefore significantly less accurate than that for \astrobj{Mkn~421} and \astrobj{Mkn~501}.  For this blazar available data from CAT (Djannati-Ata\"i et al. 2002), Whipple (Horan et al. 2002) and HEGRA (Aharonian et al. 2003) exist. These data must be distinguished in two sets.
The first set includes the HEGRA data taken in 1999-2000, CAT data taken in 1998-2000 and Whipple data taken in 2001. The second set is represented by HEGRA data taken 
in 2002 and it is characterized by a much lower flux level than for the
previous campaigns. 
\astrobj{Mkn~421} and \astrobj{Mkn~501} are maybe the best observed blazars in the high energy 
gamma-ray band and are characterized by nearly equal redshifts 
($z=0.031$ and $z=0.034$, respectively), considerably lower than 
that of \astrobj{H~1426+428}. Because of this, differences in their 
spectra (for example in the cut-off energies) cannot be explained by
different amounts of 
intervening absorption, but as due to intrinsic spectral characteristics 
(Aharonian et al. 2002b; Konopelko et al. 2003). The spectrum of \astrobj{Mkn~501} was measured up to 22 TeV both by HEGRA (Aharonian et al. 1999a, 2001a) and
by Whipple (Samuelson et al. 1998; Krennrich et al. 1999): the observations of 
these two Cherenkov telescopes are in good agreement. \astrobj{Mkn~421} was also measured up 
to 20 TeV both by HEGRA (Aharonian et al. 1999b, 2002b) and
Whipple (Krennrich et al. 1999, 2002), yielding similar
fluxes. 
HESS spectra have been collected for \astrobj{PKS~2155-304} ($z=0.117$), \astrobj{1ES1101-232} ($z=0.186$) and \astrobj{H~2356-309} ($z=0.165$) (Aharonian et al. 2005a, 2005b).
In Sec. 5 we will present spectral fits obtained for these blazars 
through the photon-photon absorption calculation, dwelling mainly on \astrobj{H~1426+428}.

\section{Results}    
As discussed in section 3.3, we have considered several models of optical and near infrared background (see  Tab. 1). 
For all these models we have calculated the optical depth due to photon-photon absorption using eq. \ref{eq:eq4}, and applied it to the theoretical 
unabsorbed blazar spectrum (eq. \ref{eq:eq6}), changing the spectral index and the normalization until the best fit was found through the $\chi{}^2$ method. Let us consider now the results obtained for each different model.


\begin{table*}
\begin{center}
\caption{$\chi^2$ values  and best fit spectral indexes ($\alpha{}$) of the considered models.  
}
\begin{tabular}{llllll}
\hline
\hline
             & C1 & MK1   & MK2   & DW1  & DW2    \\
\hline
 $\chi{}^2$  & 8.66 & 17.88 & 19.86 & 6.43 & 7.57 \\ 
 $\alpha{}$  & 2.65 & 1.60 & 1.70 & 1.80 & 1.90 \\ 
\hline
\end{tabular}
\begin{flushleft}
{\footnotesize Statistical analysis based on 12 observational data of the spectrum of \astrobj{H~1426+428} (reported in Fig. 7) with 2 parameters (the spectral index $\alpha{}$ and a normalization factor).}
\end{flushleft}
\label{tab_2}
\end{center}
\end{table*}





\subsection{Galaxy counts only}
\begin{figure}
\center{{
\epsfig{figure=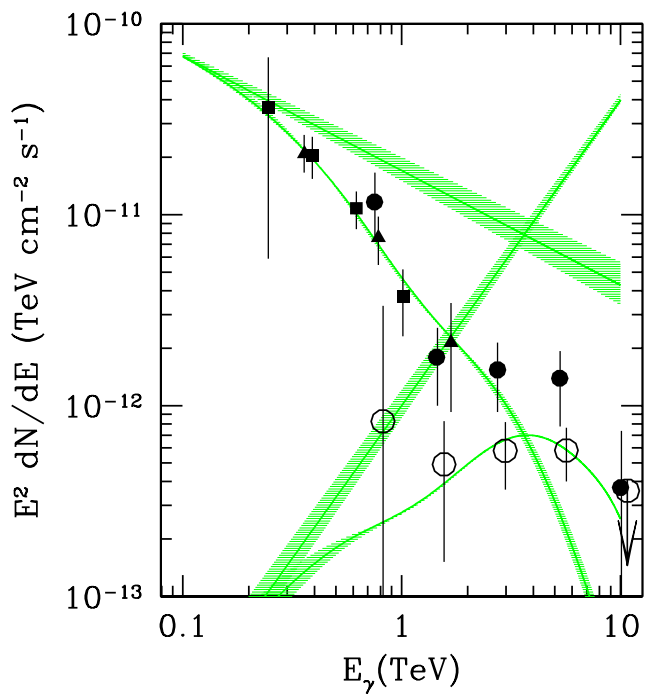,height=10cm}
}}
\caption{\label{fig:fig4}The {\it solid line} represents the best fit to the spectrum of \astrobj{H~1426+428} obtained for the model C1 (see Table~1). 
The shaded area indicates the $\pm{}1\sigma{}$ uncertainty introduced by the error in galaxy counts on the absorbed spectrum. The observational data reported here are from CAT  1998-2000 ({\it filled squares}), Whipple 2001 ({\it filled triangles}), HEGRA 1999-2000 ({\it filled circles}), HEGRA 2002 ({\it open circles}). All error bars are at 1 $\sigma{}$ (Aharonian et al. 2003). }
\end{figure}

We start from the analysis of the most conservative case (C1) in which the optical and the NIRB come only from normal galaxies as derived by Madau \&{} Pozzetti (2000);
while the MIRB/FIRB are from the rescaled Totani \&{} Takeuchi model. 
Fig.~4 shows the best fit, obtained through $\chi{}^2$ minimization, to the  \astrobj{H~1426+428} data for model C1.  
The best fit is obtained for spectral index $\alpha{}=2.65$ with $\chi{}^2\sim{}9$ (Table 2), derived considering 12 observational data (CAT  1998-2000, Whipple 2001, HEGRA 1999-2000) and 2 free parameters (spectral index $\alpha{}$ and normalization factor). This spectral index seems to be uncommonly high for this blazar. In particular, for \astrobj{H~1426+428} previous literature indicates a  spectral index $\alpha{}=1.9$ (Aharonian et al. 2002a) or $\alpha{}=1.5$ (Aharonian et al. 2003). If we consider a spectral index $\alpha{}\leq{}2$, we obtain $\chi{}^2>\,{}21$, a considerably higher value.\\
Another problem of the C1 model is that the HEGRA point at E$\sim{}5\,{}$TeV is more than 2$\sigma{}$ away from the curve. Moreover, the best fit drops at E$>3\,{}$TeV, hence providing a very poor match to the observed shape of the absorbed spectrum derived from experiments, which flattens in the energy range 1-6 TeV. In conclusion, although model C1 cannot be rejected on the basis of $\chi{}^2$ analysis alone, we consider it unlikely given its poor performance in terms of spectral slope and shape. 


\subsection{Including a NIRB excess}
We next consider models including a NIRB excess in the range $1.2-4\,{}\mu{}$m as the Matsumoto et al. (2000) data with the Kelsall's subtraction of the ZL (hereafter MK), 
and the DIRBE data with the Wright subtraction of the ZL (DW); again,    we
fix the MIRB/FIRB according to the rescaled Totani \&{} Takeuchi model in both
cases. For the optical background we experimented with all available measurements 
(Madau \&{} Pozzetti 2000; Bernstein et al. 2002; Mattila 2003).\\
\begin{figure}
\center{{
\epsfig{figure=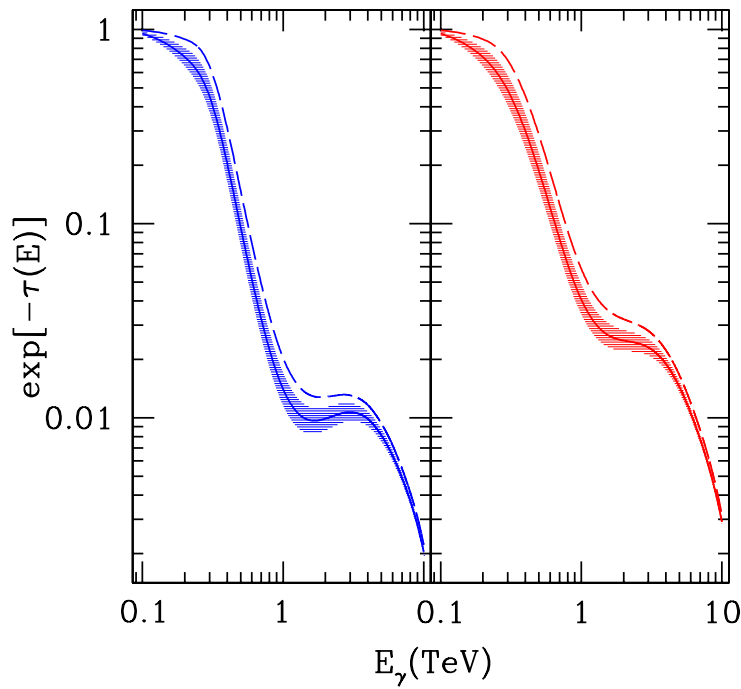,height=10cm}
}}
\caption{\label{fig:fig5} 
Optical depth for photon-photon absorption with $z_{em}=0.129$, the redshift of \astrobj{H~1426+428}. The EBL models assumed are the following. {\bf Left panel:} MK1 ({\it dashed line}) and MK2 ({\it solid}); {\bf Right panel:} DW1 ({\it dashed line}) and DW2 ({\it solid line}). Shaded areas refer to $\pm 1\sigma$ errors.}
\end{figure}

Fig.~5 shows the photon-photon absorption optical depth for a blazar at redshift 0.129 (i.e. the redshift of \astrobj{H~1426+428}), assuming the EBL models MK1+MK2 (left panel) and DW1+DW2 (right panel). The optical depth in the MK1 and MK2 cases (those which are characterized by the MK data in the NIRB) is considerably higher, especially at 1 TeV (where the absorption 
comes primarily from NIRB photons, see Fig.~1).
Fig.~6 shows the best fits for the \astrobj{H~1426+428} spectrum, again with the
NIRB from MK1 and MK2 models (left panel) and from DW1 and DW2 models (right panel). 
DW1 and DW2 provide an excellent fit, with  $\chi{}^2$ respectively 6.4 and 7.6 (Table 2). None of the observational data is more than 1$\sigma{}$ away from the fit, that correctly reproduces the shape of the absorbed spectrum suggested by the data, with the plateau at E$>1$ TeV. \\
On the contrary, the MK1 and the MK2 cases are only marginally consistent with at least 4 
Cherenkov data (CAT at $\sim{}$1 TeV and $\sim{}$0.6 TeV, HEGRA at $\sim{}$0.75 TeV  and Whipple at  $\sim{}$0.78 TeV). Also the $\chi{}^2$ is considerably higher than in the case of DW1 and DW2, even if not so high that we can reject MK1 and MK2  models.
The good result obtained adopting DW1 and DW2 models
supports the evidence for a NIRB excess, although not as pronounced as
suggested by the MK1 and MK2 models.

\begin{figure}
\center{{
\epsfig{figure=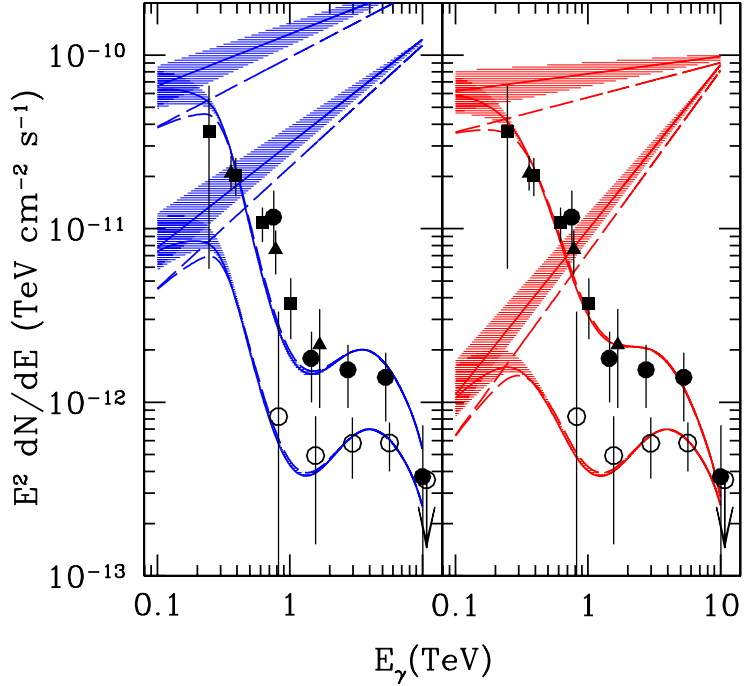,height=10cm}
}}
\caption{\label{fig:fig6} Fit to the \astrobj{H~1426+428} spectrum. The EBL models are the same as in Fig.~5; the observational data are the same as in Fig.~4. }  
\end{figure}

\subsection{Including the optical excess}        
We finally consider the differences among various measurements of the optical background. 
The left and the right panel of Fig.~6 show the differences in the spectrum of \astrobj{H~1426+428} due to the assumption of an optical background due to galaxies only or including an ``optical excess'' as measured by Bernstein et al. In both cases the fit seems to favor an optical background consistent with galaxy counts only ({\it dashed line}), although  the differences between various types of optical 
background are  smaller than the error bars in the blazar data. Thus
more solid conclusions on the optical excess have to await for more precise data.

\subsection{The spectrum of \astrobj{H~1426+428}}
From the two best models including the NIRB excess shown in Fig.~6, we can derive the shape of the intrinsic (unabsorbed) \astrobj{H~1426+428} spectrum. In both cases we find that a simple 
power-law of the form $d{N}/d{E}\propto{}E^{-\alpha{}}$ represents well the unabsorbed spectrum. In particular we find that the best fit is given by $\alpha{}=1.6$ for the MK1 model, $\alpha{}=1.7$ for MK2, $\alpha{}=1.8$ for DW1 and $\alpha{}=1.9$ for DW2. 
These values are consistent with previous works for \astrobj{H~1426+428}, which estimate a slope for the unabsorbed spectrum of this blazar of $\alpha{}=1.9$ (Aharonian et al. 2002a) or $\alpha{}=1.5$ (Aharonian et al. 2003).

\subsection{Constraints on the {\it SPITZER} measurement}
\begin{figure}
\center{{
\epsfig{figure=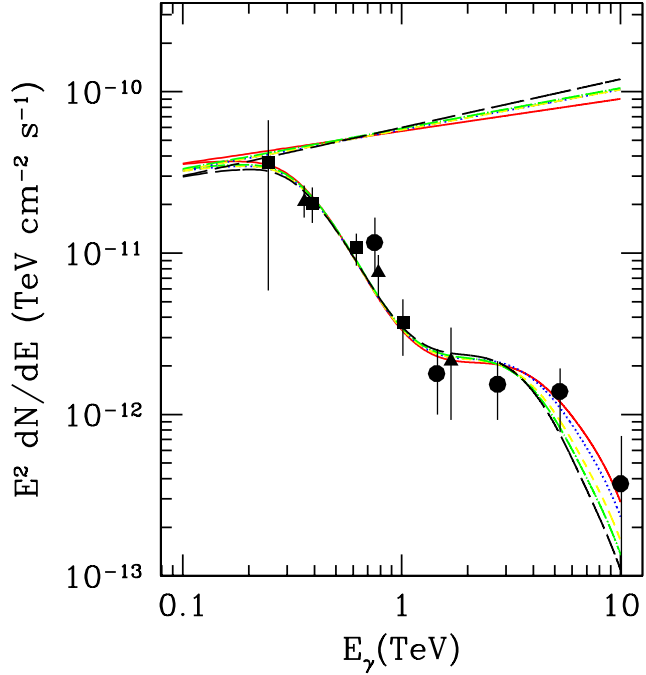,height=10cm}
}}
\caption{\label{fig:fig7} Fit of the \astrobj{H~1426+428} spectrum. We have assumed an EBL model given by: (a) optical background (0.3-1.2 $\mu{}$m): Madau \& Pozzetti 2000; (b) near infrared background (1.2-4. $\mu{}$m): DIRBE (Wright 2001) with the ZL model of Wright \& Reese (2000); (c) middle and far infrared background : model of Totani \&{} Takeuchi (2002). The model of Totani \& Takeuchi for the MIRB and the FIRB has been rescaled in the 8-30 $\mu{}$m range assuming different values of the EBL at 24 $\mu{}$m, in particular assuming: 2.7 ({\it solid line}), 3.0 ({\it dotted line}), 3.3 ({\it short dashed line}), 3.5 ({\it dot-dashed line}) and 3.8 ({\it long dashed line}) nW m$^{-2}$ sr$^{-1}$.  The observational data reported here are from CAT  1998-2000 ({\it filled squares}), Whipple 2001 ({\it filled triangles}), HEGRA 1999-2000 ({\it filled circles}).} 
\end{figure}

So far we have always used the EBL value measured by {\it SPITZER} of 2.7 nW m$^{-2}$ sr$^{-1}$ at 24 $\mu{}$m. It is interesting to check the consequences of an higher value, corresponding to the measurement upper limit of 3.8 nW m$^{-2}$ sr$^{-1}$. In Fig.~7 we assume an optical background given only by galaxy counts and a DIRBE NIRB (with the ZL model performed by Wright), which can be considered the fiducial model according to the previous analysis. For the MIRB/FIRB we adopt the Totani \&{} Takeuchi model rescaled at 24 $\mu{}$m to flux values in the range 2.7-3.8 nW m$^{-2}$ sr$^{-1}$. Fig.~7 illustrates the effect of an increase of the 24 $\mu$m flux: it becomes more and more difficult to fit the Cherenkov data, even with substantial changes of the spectral index $\alpha$. 
On the contrary, the best fit to the blazar spectrum is undoubtedly obtained adopting the {\it SPITZER} best value (2.7 nW m$^{-2}$ sr$^{-1}$).
 This result is clearly model dependent (the Totani \&{} Takeuchi model has been assumed).

\subsection{An alternative analysis}
\begin{table*}
\begin{center}
\caption{Spectral indexes ($\alpha{}$) derived using the least square fits for our models C1, DW1, MK1 and for model P1.0 of Aharonian et al. 2005b.
}
\vspace{0.2cm}
\begin{tabular}{lllll}
\hline
\hline
             & \astrobj{H~1426+428} & \astrobj{PKS~2155-304}  & \astrobj{1ES1101-232}   & \astrobj{H~2356-309}   \\
\hline
 C1  & 2.5 &  2.7 & 1.4 & 2.0 \\ 
 DW1  & 1.8 & 2.0 & 0.4 & 1.0 \\
MK1 & 1.4 & 1.6 & -0.1 & 0.1 \\
P1.0 & - & - & -0.1  & 0.7 \\
\hline
\end{tabular}
\begin{flushleft}
{\footnotesize}
\end{flushleft}
\label{tab_2}
\end{center}
\end{table*}
As discussed in Sec. 2, there is a different way to tackle the data analysis. So far we have assumed a theoretical shape (i.e. a power-law) for the unabsorbed blazar spectrum and convolved it with the optical depth for photon-photon absorption. Alternatively, one can apply $\tau(E)$ directly to the Cherenkov data by inverting eq. \ref{eq:eq6} and derive the unabsorbed blazar spectrum without making any {\it a priori} assumption about its shape. This second approach is particularly indicated when the blazar cannot be fitted by a simple power-law (for example \astrobj{Mkn~421} and \astrobj{Mkn~501}). As a sanity check, we now use this alternative approach to show that the two methods yield a coherent picture.

Fig.~8 shows the results of such attempt for \astrobj{H~1426+428}. We find that the C1 model yields 
an intrinsic spectrum which is inconsistent not only with a power-law, but also with an exponential cut-off, as a significant rise in the spectrum above 3 TeV is seen. Models MK1, and especially DW1, do not show this peculiar spectral rise at high energies. Unfortunately,  though, because of the large experimental errors in \astrobj{H~1426+428} data,  we cannot assess clearly whether the intrinsic spectrum of this blazar is better fit by a simple/broken power-law or a power-law with an exponential cut-off. 
If we assume a power-law intrinsic spectrum, we can derive the best fit spectral index with the weighted least square method. With this method we derive $\alpha{}=1.8$ for the case DW1, $\alpha{}=1.4$ for MK1 and $\alpha{}=2.5$ for C1, consistent with what we found minimizing the $\chi{}^2$ (Table~3).\\
Hence, the results of this analysis, and in particular the anomalous rise in the spectrum above 3 TeV, tend to disfavor the C1 model; the DW1 model is found again to give the best fit to the data.

\begin{figure}
\center{{
\epsfig{figure=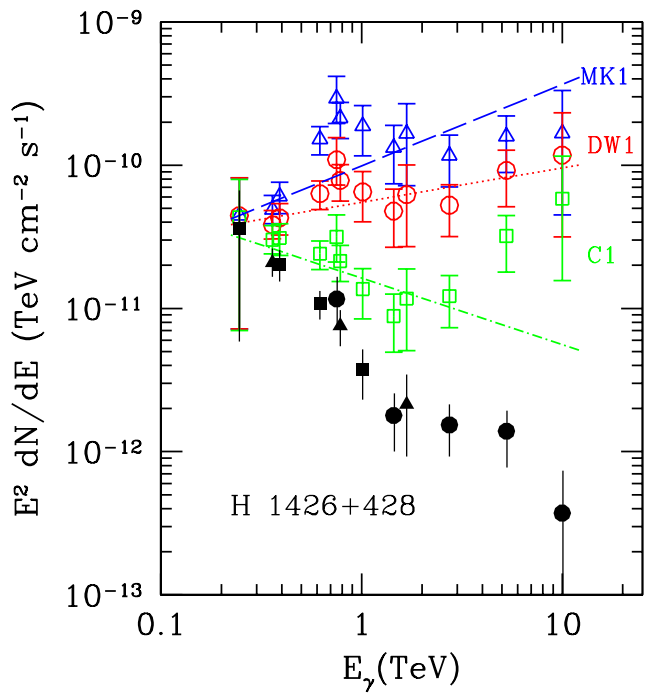,height=10cm}
}}
\caption{\label{fig:fig8} Unabsorbed spectrum of \astrobj{H~1426+428} obtained by convolving the photon-photon optical depth with the data in Fig.~7 and assuming EBL models C1 ({\it open squares}), DW1 ({\it open circles}) and MK1 ({\it open triangles}). The lines are the least square fits for the unabsorbed spectrum power-law index:  C1 ({\it dot-dashed line}, $\alpha{}=2.5$), DW1 ({\it dotted}, $\alpha{}=1.8$) and MK1 ({\it dashed}, $\alpha{}=1.4$). The error bars are obtained from the 1$\sigma$ error bars of the blazar data.} 
\end{figure}

Fig.~9 and Fig.~10 show the same procedure applied to \astrobj{Mkn~421} and \astrobj{Mkn~501}. 
The calculated optical depths for these two blazars in the range 0.7--2 TeV are in agreement with the results by Konopelko et al. (see Fig.~2 of their paper); 
above 2 TeV our results are slightly different, due to the higher MIRB flux of the EBL model  (Malkan \&{} Stecker 2001; Stecker 2003) they assume.
The data for these two blazars have significantly higher quality than those available for \astrobj{H~1426+428}; unfortunately, because of their low redshift, 
the effect of photon-photon absorption on the spectrum of these two blazars is too weak to provide additional 
constraints on the EBL. 

Figs. 11-13 show the results of this procedure applied to the new data available for \astrobj{PKS~2155-304}, \astrobj{1ES1101-232} and \astrobj{H~2356-309}, respectively. 
In the case of \astrobj{PKS~2155-304}  both the spectrum derived from the DW1 model and that derived from the MK1 model show a peculiar peak at energy $\sim{}1$ TeV. This peak is mainly due to the fact that these EBL models are characterized by a high flux at 1-4 $\mu$m and a rapid decrease at wavelengths $ >4\,{}\mu{}$m. This finding is at odd with the conclusions from the analysis of \astrobj{H~1426+428}. This is even more puzzling if we note that the two blazars are nearly at the same redshift. Instead, the unabsorbed spectrum derived from the model C1 is quite smooth.

Dwek, Krennrich \& Arendt (2005) use the theoretical emission models of Chiappetti et al. (1999), based on the EGRET data (Vestrand, Stacy \& Sreekumar 1995), to put a constraint on the  hardness of the spectrum of \astrobj{PKS~2155-304}. We have compared the results of our models with EGRET data and Chiappetti et al. models (Fig. 11). In agreement with Dwek et al. (2005), we find that the model C1 is perfectly consistent with Chiappetti et al. (1999) model. It is outside the scope of the present paper to discuss the details of synchrotron self-Compton models. Nevertheless, taking into account the errors on the least 
square fits ($\simgt 0.8$ for both DW1 and MK1 models),  on the EGRET measurements and the uncertainties of the Chiappetti et al. model, we find that the DW1 model is close to be acceptable, whereas the MK1 model has, probably, to be rejected. 
On the other hand, we also have to take into account that EGRET data and observations in the TeV range are separated by a considerable time interval (about 10 years); this can be crucial, given the variability of blazar spectra (Ghisellini, personal communication). Thus, the comparison between EGRET and TeV data, although interesting in principle, must be taken with some care.

In Fig.~12 the results for the blazar \astrobj{1ES1101-232} are presented. Aharonian et al. (2005b) have shown that the EBL models with a high excess in the NIRB must be rejected, since they imply an unabsorbed spectrum so hard that it would be difficult to explain it  within the standard hadronic or leptonic scenario (Aharonian 2001b). This conclusion is supported by our study: our model MK1 gives a spectral index $\alpha=-0.1$ (Table 3), which is too hard to be allowed by current theoretical models. However, for the model DW1 we obtain a spectral index $\alpha=0.4$ (Table 3), and for model DW2, in which the optical EBL matches the data by Bernstein et al. (2002), we get $\alpha{}=0.5$. Taking into account that the error on the spectral index from the least square  fits is large ( $\simgt 0.3$ for all the considered models),  we conclude that the unabsorbed spectra derived both from DW1 and DW2 models are hard, but consistent with the lower limit $\alpha{}=0.6$ predicted for \astrobj{1ES1101-232} by synchrotron self-Compton models (Ghisellini, personal communication). 
Then, EBL models with Wright-subtracted-ZL NIRB excess cannot be rejected on the basis of the HESS data of \astrobj{1ES1101-232}.
In Fig.~12 we show, for comparison, the model P1.0 of Aharonian et al. (2005b). This model matches the COBE/DIRBE measurements with the Wright model 
of ZL subtraction, as the DW1 model does, but it has a higher flux from 4 to 10 $\mu{}$m (Fig.~3) with respect to the latter. It is important to note
that the model P1.0 gives  a spectral index $\alpha=-0.1$, which is as hard as predicted by MK1. This points out the importance of the spectral region
4-10 $\mu{}$m to understand the EBL.

Finally, Fig.~13 shows the blazar \astrobj{H~2356-309}. Also in this case, the spectral index for the MK1 model is very hard ($\alpha{}=0.1$, Table 3), whereas both C1 and DW1 models give unabsorbed spectra of acceptable hardness.

\begin{figure}
\center{{
\epsfig{figure=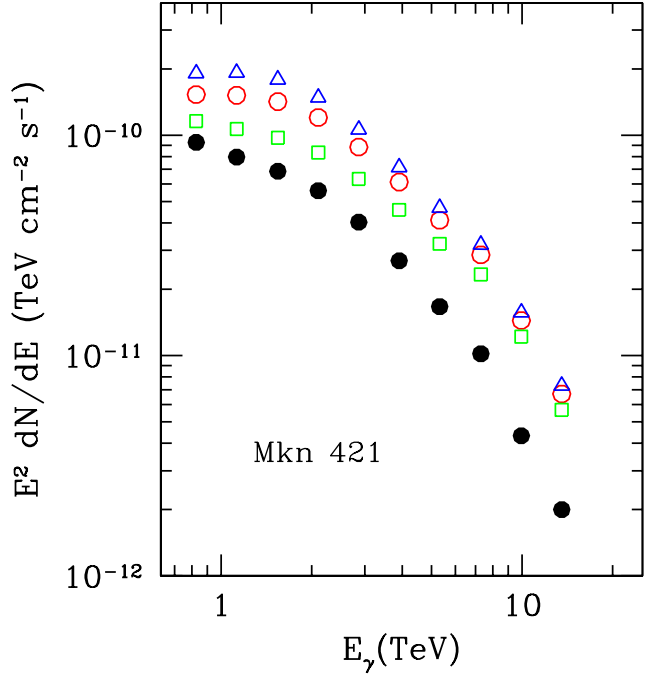,height=10cm}
}}
\caption{\label{fig:fig9} Unabsorbed spectrum of \astrobj{Mkn~421} obtained by applying the photon-photon absorption to the 2000/2001 HEGRA data (Aharonian et al. 2002b). The adopted EBL models are as in Fig.~8.}
\end{figure}

\begin{figure}
\center{{
\epsfig{figure=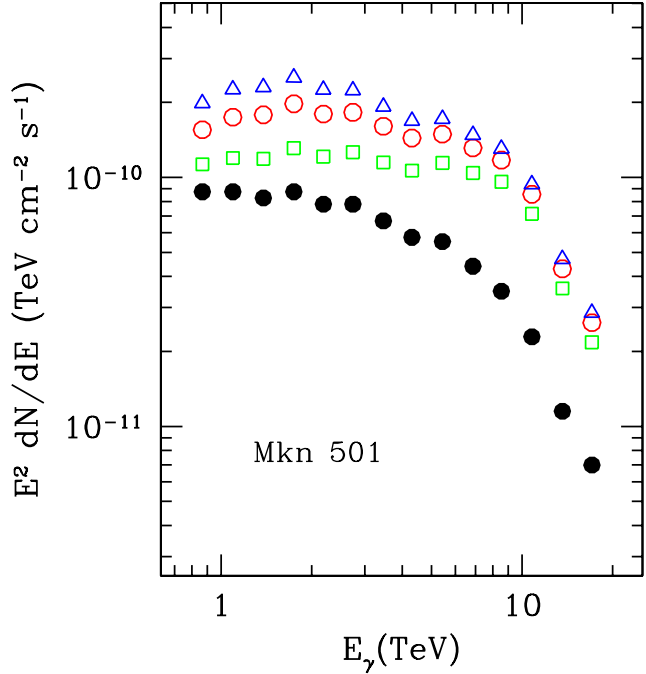,height=10cm}
}}
\caption{\label{fig:fig10} Unabsorbed spectrum of \astrobj{Mkn~501} obtained by applying the photon-photon absorption to the 1997 HEGRA data (Aharonian et al. 2002b; {\it filled squares}). }
\end{figure}

\begin{figure}
\center{{
\epsfig{figure=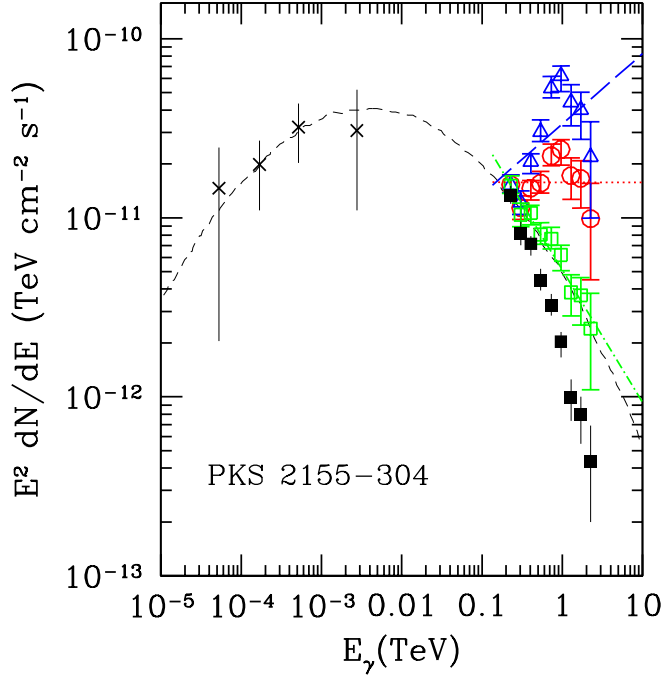,height=10cm}
}}
\caption{\label{fig:fig11} Unabsorbed spectrum of \astrobj{PKS~2155-304} obtained by applying the photon-photon absorption to the HESS data (Aharonian et al. 2005a). The adopted EBL models are as in Fig.~8. The lines are the least square fits for the unabsorbed spectrum power-law index:  C1 ({\it dot-dashed line}, $\alpha{}=2.7$), DW1 ({\it dotted}, $\alpha{}=2.0$) and MK1 ({\it dashed}, $\alpha{}=1.6$). The error bars are obtained from the 1$\sigma$ error bars of the blazar data. The {\it crosses} are the EGRET data (Vestrand, Stacy \& Sreekumar 1995) and the {\it thin short dashed line} represents the best fit synchrotron self-Compton model of Chiappetti et al. (1999).}
\end{figure}

\begin{figure}
\center{{
\epsfig{figure=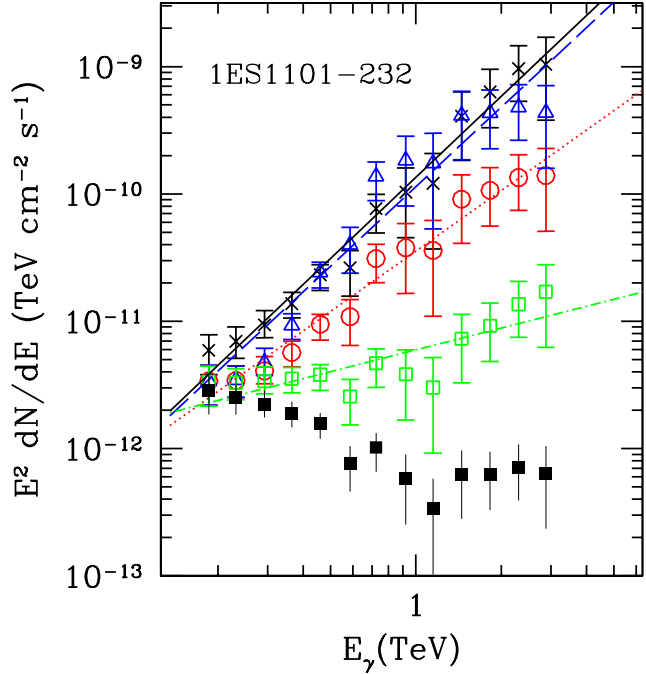,height=10cm}
}}
\caption{\label{fig:fig12} Unabsorbed spectrum of \astrobj{1ES1101-232} obtained by applying the photon-photon absorption to the HESS data (Aharonian et al. 2005b). The adopted EBL models are:  C1 ({\it open squares}), DW1 ({\it open circles}) and MK1 ({\it open triangles}). In addition we show also the results obtained for the P1.0 model of Aharonian et al. 2005b ({\it crosses}). The lines are the least square fits for the unabsorbed spectrum power-law index:  C1 ({\it dot-dashed line}, $\alpha{}=1.4$), DW1 ({\it dotted}, $\alpha{}=0.4$), MK1 ({\it dashed}, $\alpha{}=-0.1$) and P1.0 ({\it solid}, $\alpha{}=-0.1$). The error bars are obtained from the 1$\sigma$ error bars of the blazar data.}
\end{figure}
\begin{figure}
\center{{
\epsfig{figure=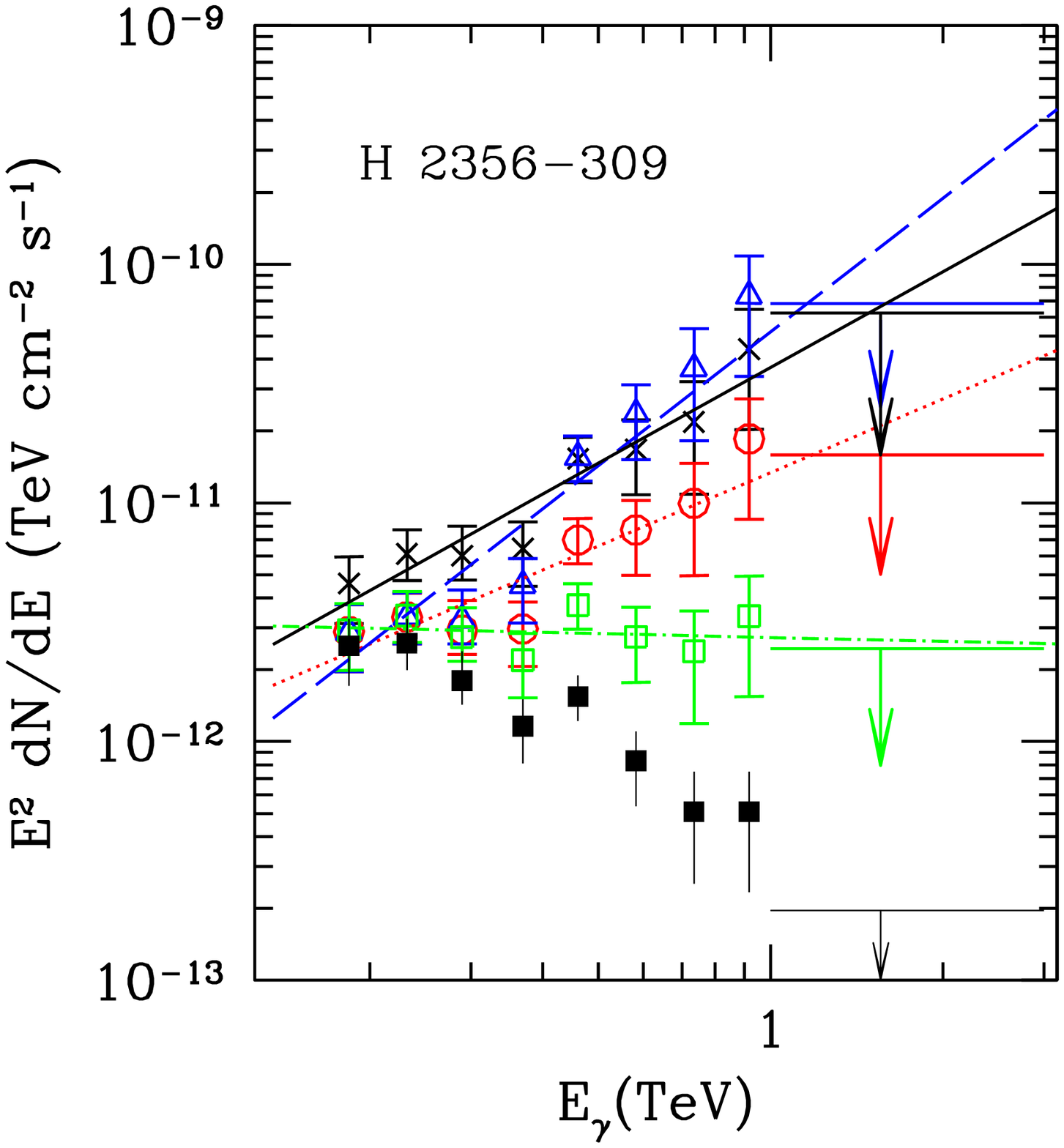,height=10cm}
}}
\caption{\label{fig:fig13} Unabsorbed spectrum of \astrobj{H~2356-309} obtained by applying the photon-photon absorption to the HESS data (Aharonian et al. 2005b). The adopted EBL models are the same as Fig.~12. The lines are the least square fits for the unabsorbed spectrum power-law index:  C1 ({\it dot-dashed line}, $\alpha{}=2.0$), DW1 ({\it dotted}, $\alpha{}=1.0$), MK1 ({\it dashed}, $\alpha{}=0.1$) and P1.0 ({\it solid}, $\alpha{}=0.7$). The error bars are obtained from the 1$\sigma$ error bars of the blazar data.}
\end{figure}

\section{Conclusions}
In this paper, we have explored the effects produced by photon-photon absorption on blazar 
spectra to put constraints on extragalactic background light from the optical to the far infrared bands. We mostly focused on the high redshift blazar \astrobj{H~1426+428}, for which we adopted a power-law unabsorbed spectrum. This might be a simplifying assumption, as some theoretical models indicate a more complex spectral shape (Inoue and Takahara 1996; Tavecchio, Maraschi \& Ghisellini 1998; Fossati et al. 2000). On the other hand, the data for \astrobj{H~1426+428} are not sufficiently accurate to suggest the existence of a cut-off or to exclude a power-law spectrum. 

We have considered three different models of the NIRB, based respectively on galaxy counts (C1) and on the presence of an excess (DW and MK). In the case of \astrobj{H~1426+428} the DW model provides the best fit both using the method presented in Sec. 5.2, and the alternative one of Sec. 5.6. The MK model has a higher $\chi{}^2$ but cannot be rejected. For the same reasons, C1 model cannot be excluded either; however, in this case a very peculiar blazar intrinsic spectrum must be assumed (Fig. 8). We conclude that the presence of a NIRB excess with respect to galaxy counts, at the level given by the DW model, seems to be required to fit the blazar spectrum.

The very recent HESS data of the blazars \astrobj{PKS~2155-304} (Aharonian et al. 2005a), \astrobj{1ES1101-232} and \astrobj{H~2356-309} (Aharonian et al. 2005b) allow us to put stronger constraints to the NIRB. In particular, from the analysis of all these blazar, we conclude that the MK models, based on the Kelsall method of ZL subtraction, must be rejected, since they imply too hard blazar spectra. The spectra of \astrobj{PKS~2155-304} and \astrobj{1ES1101-232} are marginally consistent with DW models, based on Wright ZL subtraction, which can be considered as an upper limit to the NIRB. The model C1, without NIRB excess, is favored by all these new data.
These findings are quite different from those we derived for \astrobj{H~1426+428}. Since the available data for  \astrobj{H~1426+428} are old, new measurement of this blazar are eagerly required, to shed some light on this puzzling inconsistency.

The derived constraints on the optical EBL are weaker, due to the fact that deviations between different optical EBL models are comparable to the experimental errors. The fit tends to favor models without an optical excess over galactic light, contrary to the result obtained by  Bernstein et al. (2002). A more solid conclusion on the amplitude of the optical EBL has to await for the next generation of Cherenkov Telescopes as {\it MAGIC, VERITAS, HESS, GLAST}, or infrared satellites ({\it CIBER}). 

Finally, in the mid-infrared the {\it SPITZER} measurement of $\nu{}I_{\nu{}}$=2.7 nW m$^{-2}$ sr$^{-1}$ at 24 $\mu{}$m, combined with the EBL model by Totani \& Takeuchi (2002), allows us to obtain a good fit for all the blazars available. 
 Again, a tremendous advance on the determination of the MIRB/FIRB is expected from   {\it SPITZER} and the next generation of infrared satellites as {\it ASTRO-F}, and {\it SPICA}.\\


In summary, recent measurements of blazars in the TeV range seem to exclude the existence of a strong NIRB excess consistent with Kelsall's model of ZL subtraction. The COBE/DIRBE measurements, after Wright's model ZL subtraction, represent a firm NIRB upper limit. 

\section*{Acknowledgements}
We thank G.~Ghisellini  for useful discussions, and acknowledge J.~Bullock, L.~Metcalfe, A.~Celotti, L.~Costamante, S.~Inoue, T.~Matsumoto, K.~Mattila, J. Primack and F.~W.~Stecker 
for enlightening comments. This work was partially supported by the Research and Training Network `The Physics of the Intergalactic Medium' set up by the European Community
under the contract HPRN-CT2000-00126 RG29185.


\end{document}